\documentclass[usenatbib]{mn2e}

\usepackage{amsmath, mathrsfs, amssymb}
\usepackage{graphicx}
\usepackage{rotating}
\usepackage{longtable,lscape}
\usepackage{morefloats}
\usepackage{color}
\usepackage[usenames,dvipsnames,svgnames,table]{xcolor}
\usepackage{enumitem}
\usepackage{caption}
\usepackage{verbatim}
\usepackage{color}

\newcommand{\del}{\partial}

\title[Strongly magnetized discs]{Strongly magnetized accretion discs require poloidal flux}

\author[Salvesen et al.]{Greg~Salvesen$^{1,2}\thanks{E-mail: salvesen@colorado.edu}\thanks{NASA Earth and Space Science Graduate Fellow.}$, Philip~J.~Armitage$^{1,2}$, Jacob~B.~Simon$^{1,3}$\thanks{Sagan Fellow.}, \& \newauthor Mitchell~C.~Begelman$^{1,2}$ \\
$^{1}${JILA, University of Colorado and National Institute of Standards and Technology, 440 UCB, Boulder, CO 80309-0440, USA.} \\
$^{2}${Department of Astrophysical and Planetary Sciences, University of Colorado, 391 UCB, Boulder, CO 80309-0391, USA.} \\
$^{3}${Department of Space Studies, Southwest Research Institute, Boulder, CO 80302, USA.}}

\begin{document}
\label{firstpage}
\maketitle

\begin{abstract}
Motivated by indirect observational evidence for strongly magnetized accretion discs around black holes, and the novel theoretical properties of such solutions, we investigate how a strong magnetization state can develop and persist.  To this end, we perform local simulations of accretion discs with an initially purely toroidal magnetic field of equipartition strength.  We demonstrate that discs with zero net vertical magnetic flux and realistic boundary conditions cannot sustain a strong toroidal field.  However, a magnetic pressure-dominated disc can form from an initial configuration with a sufficient amount of net vertical flux and realistic boundary conditions.  Our results suggest that poloidal flux is a necessary prerequisite for the sustainability of strongly magnetized accretion discs.
\end{abstract}

\begin{keywords}
accretion, accretion discs - dynamo - instabilities - magnetohydrodynamics (MHD) - turbulence - X-rays: binaries
\end{keywords}

\section{Introduction}
\label{sec:intro}
Magnetic fields are fundamental to the physics of accretion discs.  The magnetorotational instability (MRI) causes an accretion disc to enter a self-sustaining turbulent steady state, wherein an effective viscosity is generated that drives accretion \citep{BalbusHawley1991, BalbusHawley1998}.  The disc magnetization is parametrized by the ratio of gas-to-magnetic pressure, $\beta \equiv p_{\rm gas} / p_{B}$.  While some numerical simulations examine MRI turbulence in strongly magnetized discs \citep[$\beta \lesssim 1$; e.g.,][]{JohansenLevin2008, BaiStone2013, Salvesen2016a}, the majority focus on the weakly magnetized regime.

 Several lines of evidence suggest that real accretion discs around black holes in X-ray binaries and galactic nuclei may  be strongly magnetized.  Observations of powerful disc winds \citep[e.g.,][]{Miller2006a} and relativistic jets \citep[e.g.,][]{Fender2004} in black hole X-ray binary systems suggest the presence of a significant poloidal flux, which catalyzes a much stronger toroidal field in MRI-active discs. On larger scales, the gas supplied to the disc by the donor star is expected to be strongly magnetized \citep[$\beta \lesssim 1$;][]{JohansenLevin2008}. In the Galactic Center, dust polarization measurements in the inner $\sim 10^2 \ {\rm pc}$ reveal both a toroidal magnetic field near the Galactic plane, and a large-scale poloidal field at higher Galactic latitudes \citep{Nishiyama2010}.

At a more circumstantial level, strongly magnetized discs provide a promising theoretical framework for addressing longstanding problems that plague their weakly magnetized counterparts. In particular, magnetically dominated discs are not prone to thermal instability in radiation pressure dominated regions \citep{BegelmanPringle2007}, a result which appears consistent with the low levels of variability seen in luminous X-ray binary states. They are also less susceptible to gravitational fragmentation that would inhibit continued accretion onto the black hole in active galactic nuclei \citep{Pariev2003, BegelmanPringle2007, Gaburov2012}.  Furthermore, the effective $\alpha$-viscosity \citep{ShakuraSunyaev1973}, which parametrizes the rate of angular momentum transport, is observationally constrained to be $\alpha \sim 0.1-0.4$ in dwarf novae systems \citep{King2007}. Local disc simulations with weak magnetic pressure support and zero net vertical flux find $\alpha \sim 0.01$ \citep{Stone1996, Davis2010, Simon2011} .  However, local disc simulations with net vertical flux allow the disc to develop significant magnetic pressure support and boost $\alpha$ to values $\sim 0.1-1$ \citep[e.g.,][]{Hawley1995, Salvesen2016a}.  Finally, recent theoretical models for X-ray binary state transitions appeal to strongly magnetized accretion discs \citep{Begelman2015} and the evolution of a net poloidal flux \citep{BegelmanArmitage2014}.

In this Letter, we address the question of whether there are multiple routes by which an accretion disc can sustain a strongly magnetized state. One well-studied route requires a net poloidal flux (typically in the window $10^3 > \beta_p > 10$) that is strong enough to affect the MRI dynamics but weak enough to admit MRI-driven turbulence. Simulations show that this leads to a supra-thermal mid-plane field that is replenished by dynamo action as it buoyantly escapes \citep{BaiStone2013, Salvesen2016a}.  An alternate route could involve processes other than the MRI.  In particular, \citet{JohansenLevin2008} argued that a dynamo mediated by the Parker instability could maintain a strongly magnetized state from an initial configuration of a purely toroidal field in equipartition with the gas. This suggestion is particularly interesting as --- if correct --- it would imply that discs with zero net poloidal flux could exist in two distinct states depending upon the history of the system.

Here we consider local disc simulations that have zero net vertical magnetic flux and are initialized with an equipartition toroidal field. We adopt two sets of vertical boundary conditions, one which traps toroidal magnetic flux and one which allows it to escape freely. We show that the disc cannot maintain its strongly magnetized initial state for the more physically realistic outflow boundary conditions, and hence conclude that strongly magnetized astrophysical accretion discs require a background poloidal magnetic flux.

\section{Numerical Simulations}
\label{sec:sims}
Following \citet{Salvesen2016a}, we use the \texttt{Athena} code \citep{GardinerStone2005, GardinerStone2008, Stone2008} to solve the equations of compressible, isothermal, ideal magnetohydrodynamics in the ``shearing box'' approximation \citep{Hawley1995, StoneGardiner2010} including vertical density stratification,
\begin{align}
\frac{\del{\rho}}{\del{t}} = &- \mathbf{\nabla} \cdot \left( \rho \mathbf{v} \right) \label{eqn:mass} \\
\frac{\del{\left( \rho \mathbf{v} \right)}}{\del{t}} = &- \mathbf{\nabla} \cdot \left[ \rho \mathbf{v} \mathbf{v} - \mathbf{B} \mathbf{B} + \left( p_{\rm gas} + \frac{B^{2}}{2} \right) \mathbf{I} \right] \nonumber \\
&+ 2 q \rho \Omega^{2} x \mathbf{\hat{i}} - \rho \Omega^{2} z \mathbf{\hat{k}} - 2 \mathbf{\Omega} \times \left( \rho \mathbf{v} \right) \label{eqn:momentum} \\
\frac{\del{\mathbf{B}}}{\del{t}} = &- \mathbf{\nabla} \cdot \left( \mathbf{v} \mathbf{B} - \mathbf{B} \mathbf{v}\right). \label{eqn:magflux}
\end{align}
In order, Equations \ref{eqn:mass} - \ref{eqn:magflux} describe the conservation of mass, the conservation of momentum, and magnetic induction, with the various parameters having their usual meanings: $\rho$ is the gas density, $p_{\rm gas}$ is the gas pressure, $\mathbf{v}$ is the gas velocity, and $\mathbf{B}$ is the magnetic field. Our magnetic field definition subsumes a factor of $\mu / \sqrt{4 \pi}$, with $\mu = 1$ being the magnetic permeability.  We choose an equation of state, $p_{\rm gas} = \rho c_{\rm s}^{2}$, corresponding to an isothermal gas with sound speed $c_{\rm s}$.  $\mathbf{I}$ is the identity matrix and $(\mathbf{\hat{i}}, \mathbf{\hat{j}}, \mathbf{\hat{k}})$ are the Cartesian $\left( x, y, z \right)$ unit vectors.  Specific to the shearing box geometry, $\mathbf{\Omega} = \Omega \mathbf{\hat{k}}$ is the angular frequency corresponding to co-rotation with the disc.  The simulation domain is centered on the arbitrary reference radial location $R = R_{0}$ and $q = - {\rm d} {\rm ln}\left( \Omega \right) / {\rm d} {\rm ln}\left( R \right) = 3/2$ is the shear parameter corresponding to Keplerian rotation.

The question we are interested in answering is whether a background poloidal magnetic flux is necessary to sustain a strongly magnetized accretion disc.  The net vertical magnetic flux simulations of \citet{Salvesen2016a} with $\beta_{0}^{\rm mid} \lesssim 10^{3}$ and outflow boundary conditions developed into a magnetic pressure-dominated state.  The zero net vertical flux simulations of \citet{JohansenLevin2008} were initialized with an equipartition toroidal field and remained strongly magnetized, but adopted boundary conditions that confined the magnetic flux within the domain.  The simulations presented in this Letter (see Table \ref{tab:sims}) are very similar to those of \citet{Salvesen2016a}, but have the following two differences: (1) initial conditions with a purely toroidal magnetic field configuration (\S \ref{sec:ICs}) and (2) boundary conditions that either do or do not permit magnetic flux to escape (\S \ref{sec:BCs}).

\subsection{Initial Conditions}
\label{sec:ICs}
As did \citet{JohansenLevin2008}, we initialize zero net vertical magnetic flux simulations with a vertical gas density profile in hydrostatic equilibrium and a purely toroidal magnetic field configuration,
\begin{align}
\rho\left( x, y, z \right) &= \rho_{0}^{\rm mid} {\rm exp}\left( \frac{- z^{2}}{2 H_{\beta}^{2}} \right) \\
B_{y, 0}\left( x, y, z \right) &= \left[ 2 \beta_{0}^{-1} c_{\rm s}^{2} \rho_{0}^{\rm mid} {\rm exp}\left( \frac{- z^{2}}{2 H_{\beta}^{2}} \right) \right]^{1/2},
\end{align}
with $\rho_{0}^{\rm mid} = 1$. The gas scale height is $H_{\beta} = \sqrt{1 + \beta_{0}^{-1}} c_{\rm s} / \Omega$.  Because we choose $c_{\rm s} = 1$ and $\Omega = 1$, the scale height resulting from thermal pressure support alone opposing vertical gravity is $H_{0} = c_{\rm s} / \Omega = 1$, which we adopt as our unit of length.  The domain size for all simulations is $\left( L_{x}, L_{y}, L_{z} \right) = \left( 10H_{0}, 20H_{0}, 10H_{0} \right)$.  Disc magnetization is parametrized by the ratio of gas-to-magnetic pressure,
\begin{equation}
\beta \equiv \frac{p_{\rm gas}}{p_{B}} = \frac{\rho c_{\rm s}^{2}}{B^{2} / 2}. \label{eqn:beta}
\end{equation}
We choose an equipartition initial magnetic field $\beta_{0} = 1$, which makes the vertical domain size $7.1H_{\beta}$ when the magnetic field contribution to hydrostatic equilibrium is included.  Random perturbations to the gas density and velocity are introduced at $t = 0$ in order to initiate the MRI \citep{Hawley1995}. We impose a density floor $\rho_{\rm floor} = 10^{-4} \rho_{0}^{\rm mid}$.

\begin{table*}
\addtolength{\tabcolsep}{-1pt}
\centering
\begin{tabular}{c c c c c c c c c c c c}
\hline
\hline
ID & Grid Res. & $\beta_{0}^{\rm mid}$ & $B_{z}$ & Vertical BCs & $t_{\rm f}$ & $\langle Q_{y}^{\rm mid} \rangle_{t}$ & $\langle Q_{z}^{\rm mid} \rangle_{t}$ & $\langle T_{xy, {\rm Rey}}^{\rm mid} \rangle_{t}$ & $\langle T_{xy, {\rm Max}}^{\rm mid} \rangle_{t}$ & $\langle \widehat{\alpha}^{\rm mid} \rangle_{t}$ & $\langle \widehat{\beta}^{\rm mid} \rangle_{t}$ \\
 & [$N_{\rm zones} / H_{0}$] & & & & [orbits] & & & & & & \\
\hline
ZNVF-O & 36 & 1 & 0 & Outflowing & 125 & 44(2) & 10.3(5) & 0.0038(5) & 0.013(1) & 0.013(1) & 45(4) \\
ZNVF-P & 24 & 1 & 0 & Periodic & 125 & 35(2) & 8.4(4) & 0.008(1) & 0.022(2) & 0.019(2) & 32(3) \\
NVF-$\beta 2{^\ast}$ & 24 & $10^{2}$ & 0.141 & Outflowing & 225 & 3.3(7)e2 & 9(1)e1 & 0.10(6) & 0.5(1) & 1.0(4) & 0.4(1) \\
\hline
\end{tabular}
\caption{Summary of simulations.  From {\it left} to {\it right} the columns are: simulation identification label, grid resolution (applies to all dimensions), initial plasma-$\beta$ at the disc mid-plane, net vertical magnetic flux density (code units), vertical boundary conditions, and simulation termination time.  All subsequent quantities, which are defined in the text, are evaluated at the disc mid-plane and time-averaged from $t_{\rm i}$ = 25 to $t_{\rm f}$.  Parentheses indicate the $\pm 1 \sigma$ range on the last digit from the time averaging.  $^{\ast}$Reference net vertical magnetic flux simulation from \citet{Salvesen2016a}.}
\label{tab:sims}
\end{table*}

\subsection{Boundary Conditions}
\label{sec:BCs}
We adopt the usual shearing box boundary conditions of shearing periodic in $x$ (radial) and strictly periodic in $y$ (toroidal) \citep{Hawley1995, Simon2011}.  Table \ref{tab:sims} summarizes our simulations, which fundamentally differ only in their choice of $z$ (vertical) boundary conditions:
\begin{itemize}
\item ZNVF-O adopts the modified outflowing vertical boundary conditions of \citet{Simon2011} and \citet{Simon2013}, in which the gas density, pressure, and in-plane components of the magnetic field are exponentially extrapolated from the physical domain into the ghost zones.  This method prevents the unphysical accumulation of magnetic flux at the vertical boundaries, thus allowing magnetic flux to escape the simulation domain.   
\item ZNVF-P adopts periodic vertical boundary conditions, which do {\it not} allow magnetic flux to escape the simulation domain.
\end{itemize}

Neither condition is fully realistic, but outflowing boundary conditions are closer to the physical situation.  Any significant differences between the zero net vertical flux simulations ZNVF-O and ZNVF-P can be ascribed to their differences in vertical boundary conditions.  We performed resolution and vertical domain size convergence studies that corroborated the results from simulation ZNVF-P.  We contrast these runs to NVF-$\beta 2$ from \citet{Salvesen2016a}, which combines outflow boundary conditions with a net poloidal field $\beta_0^{\rm mid} = 10^2$.

\section{Analysis and Results}
\label{sec:analres}
We are interested in the turbulent steady states arising from zero net vertical flux shearing box simulations that differ in their choice of vertical boundary conditions.  Empirically, the saturated state is considered well-resolved for quality factors $Q_{i} \gtrsim 6$ \citep{Sano2004}, given by\footnote{\citet{Hawley2011} suggest $Q_{y} \gtrsim 20$ and $Q_{z} \gtrsim 10$.},
\begin{equation}
Q_{i} = \frac{\lambda_{\rm MRI}}{\Delta x_{i}},
\end{equation}
where, for spatial dimension $i$, the characteristic MRI wavelength is $\lambda_{{\rm MRI}, i} = 2 \pi v_{{\rm A}, i} / \Omega$ and the Alfv\'{e}n speed is $v_{{\rm A}, i} = \sqrt{B^{2}_{i} / \rho}$.  Table \ref{tab:sims} shows that all of our simulations are well-resolved according to the $Q_{y}$ and $Q_{z}$ benchmarks.

\begin{figure}
  \includegraphics[width=84mm]{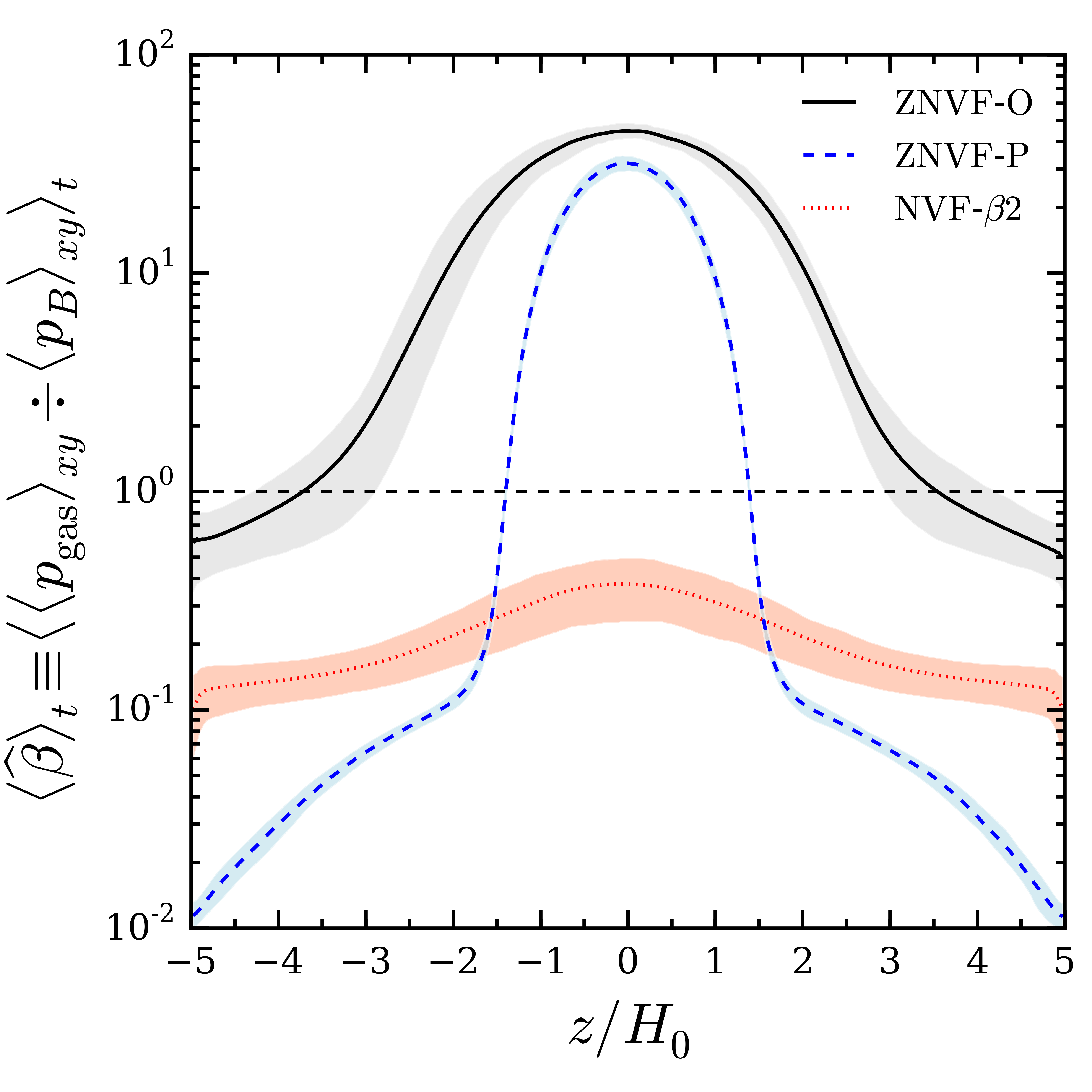}
  \caption{Vertical profiles of the time-averaged plasma-$\beta$ for zero net vertical flux simulations with outflowing ({\it solid black line}) and periodic ({\it dashed blue line}) vertical boundary conditions.  The strongly magnetized net vertical flux simulation with outflowing boundary conditions ({\it dotted red line}) from \citet{Salvesen2016a} is shown for reference.  Coloured bands show the respective $\pm 1 \sigma$ range in $\widehat{\beta}$ from the time averaging.  The {\it horizontal dashed line} marks equipartition $\widehat{\beta} = 1$.}
  \label{fig:zpro_beta}
\end{figure}

Table \ref{tab:sims} also lists diagnostics of MRI turbulence, each evaluated at the disc mid-plane and time-averaged over the saturated state (i.e., $t \ge 25$ orbits).  The rate of angular momentum transport is parametrized by the effective $\alpha$-viscosity \citep{ShakuraSunyaev1973},
\begin{equation}
\widehat{\alpha} = \frac{\langle T_{{xy}, {\rm Rey}} \rangle_{xy} + \langle T_{{xy}, {\rm Max}} \rangle_{xy}}{\langle p_{\rm gas} \rangle_{xy}},
\end{equation}
where the notation $\langle \cdot \rangle_{xy}$ indicates that the quantity is horizontally-averaged.  The Reynolds and Maxwell contributions to the $xy$ component of the total stress tensor are $T_{{xy}, {\rm Rey}} = \rho v_{x} v_{y}^{\prime}$ and $T_{{xy}, {\rm Max}} = - B_{x} B_{y}$, where $v_{y}^{\prime}$ are the remaining fluctuations when the background shear is subtracted from $v_{y}$.  The plasma-$\beta$ parameter in Table \ref{tab:sims} is defined as,
\begin{equation}
\widehat{\beta} = \frac{\langle p_{\rm gas} \rangle_{xy}}{\langle p_{B} \rangle_{xy}}.
\end{equation}
Figure \ref{fig:zpro_beta} shows the vertical profiles of $\widehat{\beta}$ for each simulation. 

To our surprise, the {\em mid-plane} diagnostics of MRI turbulence are only weakly dependent on the vertical boundary conditions for the zero net flux simulations. In the saturated turbulent steady state of runs ZNVF-O and ZNVF-P, we find $T_{{xy}, {\rm Max}}^{\rm mid} \simeq 4 T_{{xy}, {\rm Rey}}^{\rm mid}$, $\widehat{\alpha}^{\rm mid} \sim 0.01$, and $\widehat{\beta}^{\rm mid} \gg 1$, consistent with previous shearing box simulations with zero net vertical magnetic flux \cite[e.g.,][]{Stone1996,Davis2010,Simon2011}.  Conversely, the strong net vertical magnetic flux simulation NVF-$\beta 2$ shows an enhanced $\widehat{\alpha}^{\rm mid} \simeq 1.0$ and a magnetic pressure-dominated disc with $\widehat{\beta}^{\rm mid} \simeq 0.4$ \citep{Salvesen2016a}.

\begin{figure*}
  \includegraphics[width=84mm]{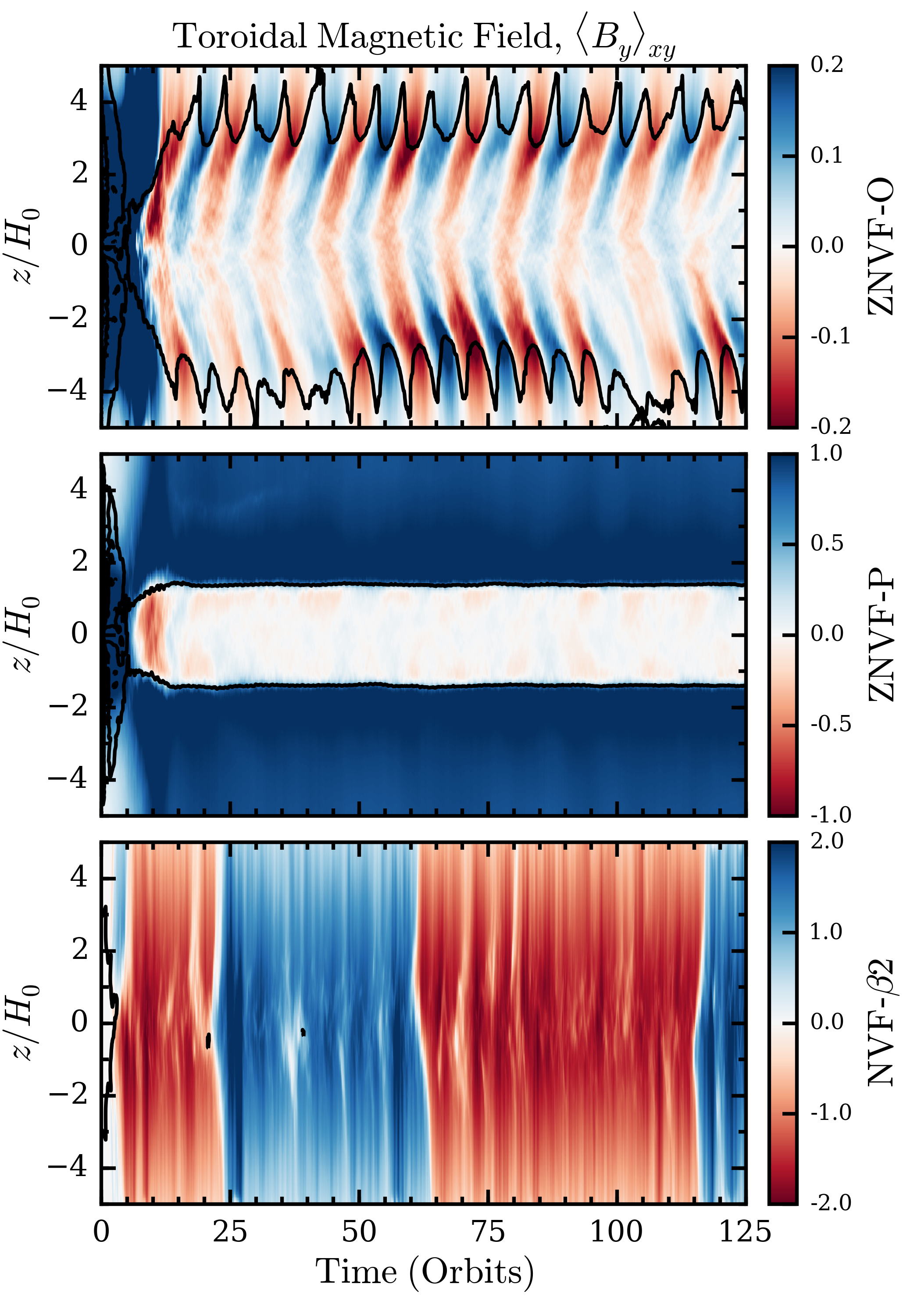}
  \includegraphics[width=84mm]{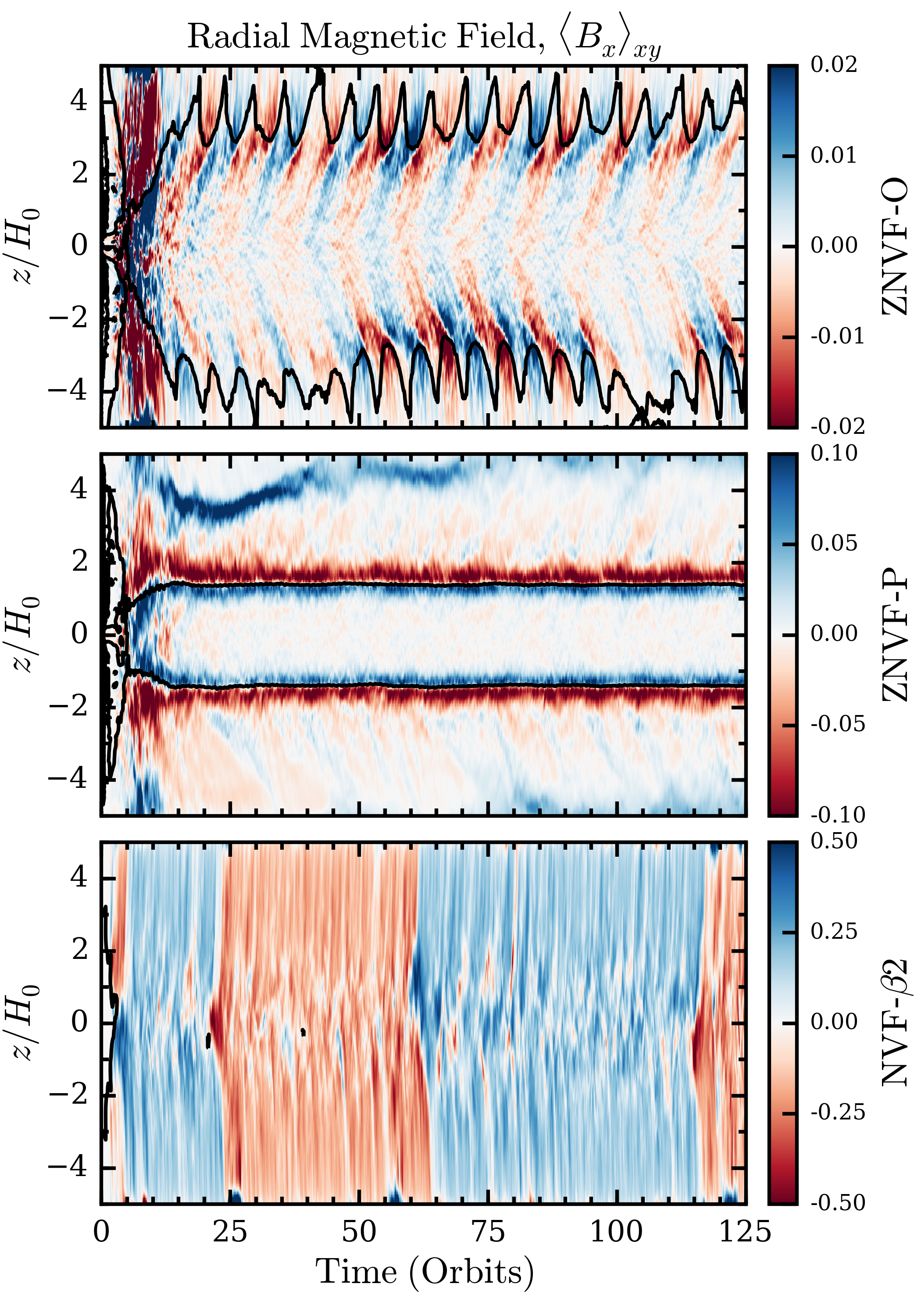}
  \caption{{\it Left panels:} Space-time diagrams of the horizontally-averaged toroidal magnetic field, $\langle B_{y} \rangle_{xy}$. {\it Right panels:} Space-time diagrams of the horizontally-averaged radial magnetic field, $\langle B_{x} \rangle_{xy}$.  Shown are simulations ZNVF-O ({\it top panels}),  ZNVF-P ({\it middle panels}),  and NVF-$\beta 2$ ({\it botom panels}).  {\it Black lines} mark the $\beta = 1$ contour, exterior (interior) to which $\beta < 1$ ($\beta > 1$).  The magnetic field reversals characteristic of the MRI-dynamo are only observed in the simulations with outflowing boundary conditions (i.e., ZNVF-O and NVF-$\beta 2$).  This suggests that simulations with vertical boundary conditions that confine magnetic flux within the domain cannot capture important accretion disc physics.}
  \label{fig:spacetime_By}
\end{figure*}

Larger differences between the zero net flux simulations are apparent from the vertical profiles of magnetization shown in Figure~\ref{fig:zpro_beta}. With outflowing boundary conditions, we find that the initially strong (equipartition level) toroidal field buoyantly escapes in the transient stage. The disc evolves into a generally weakly magnetized state, with $\widehat{\beta} \sim 1$ being reached only in a coronal region at $|z| > 3.5 H_0$.  With periodic boundary conditions, on the other hand, the final magnetized state is determined by the redistribution of the initial equipartition toroidal field. After a sufficiently long interval, this leads to a weakly magnetized disc core within about one scale height of the disc mid-plane, with extremely strong magnetization at higher altitudes.  We are only able to sustain a strongly magnetized disc mid-plane with the aid of poloidal field.  Simulation NVF-$\beta 2$, which also adopts outflowing boundary conditions, evolves from a net vertical flux configuration with $\beta_{0}^{\rm mid} = 10^{2}$ into a strongly magnetized state with $\widehat{\beta} \lesssim 1$. In this run the MRI-dynamo constantly replenishes the escaping toroidal field from the imposed vertical field.  

Figure \ref{fig:spacetime_By} ({\it left panels}) shows the space-time diagrams of the toroidal magnetic field for each simulation.  The simulations with outflowing vertical boundary conditions (ZNVF-O and NVF-$\beta 2$) exhibit the toroidal field reversals characteristic of MRI-dynamo activity, albeit with differing periods.  Simulation ZNVF-P, which confines magnetic flux within the domain, does not show this behaviour of periodically launching current sheets across the full vertical domain.  This illustrates that other important aspects of MRI-dynamo activity in accretion discs, beyond the vertical profile of magnetization, are sensitive to the vertical boundary conditions and cannot be captured by simulations that do not permit magnetic flux to buoyantly escape the domain.

Figure \ref{fig:spacetime_By} ({\it right panels}) also shows space-time diagrams for the radial component of the magnetic field.  Similarly to the toroidal field, the radial field experiences dynamo cycles of escape and replenishment for simulations ZNVF-O and NVF-$\beta2$.  In simulation ZNVF-P, the radial field develops long-lived banded structures at the $\beta \simeq 1$ interfaces ($| z / H_{0} | \simeq 1.5$).  The simulations of \citet{JohansenLevin2008} developed similar peaks in the radial field, but at larger scale heights and also at the disc mid-plane, which we do not observe in simulation ZNVF-P.

\begin{figure*}
  \includegraphics[width=56mm]{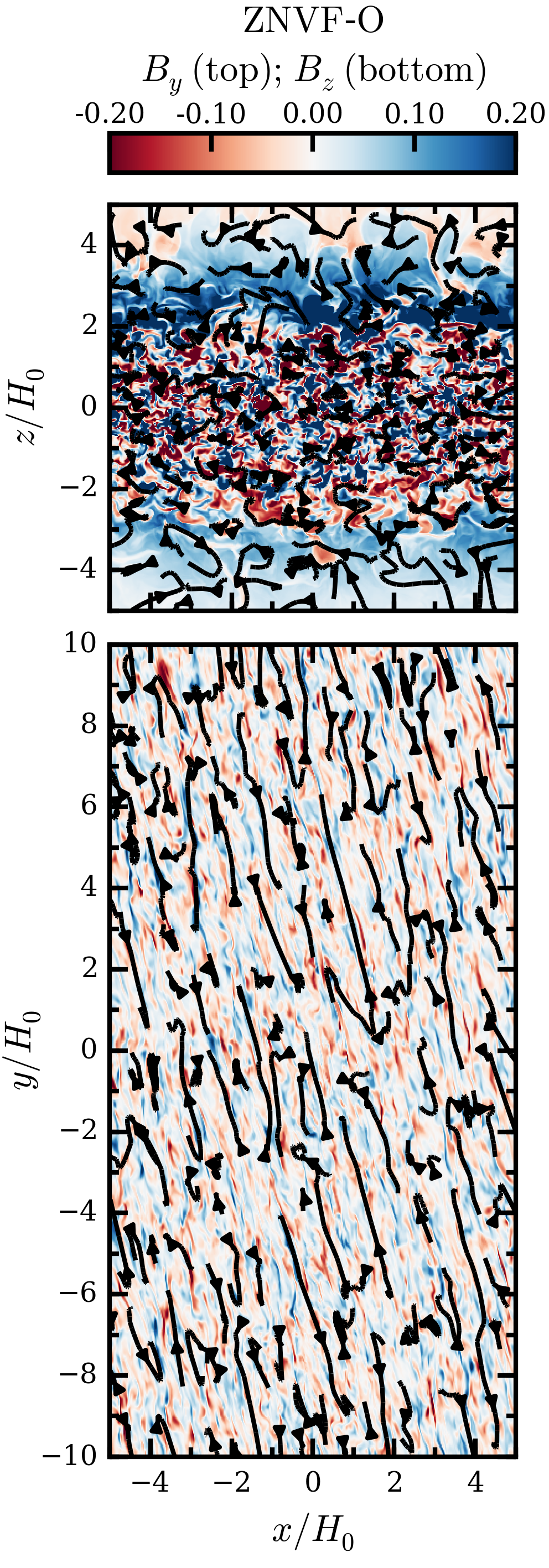}
  \includegraphics[width=56mm]{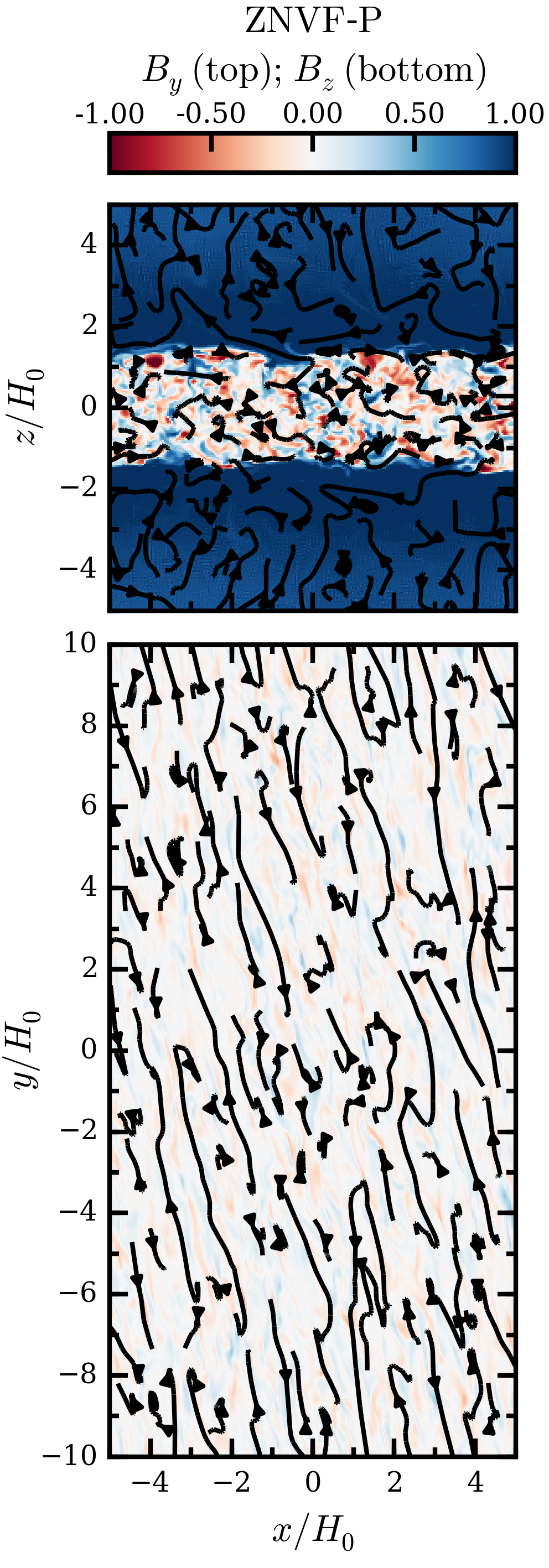}
  \includegraphics[width=56mm]{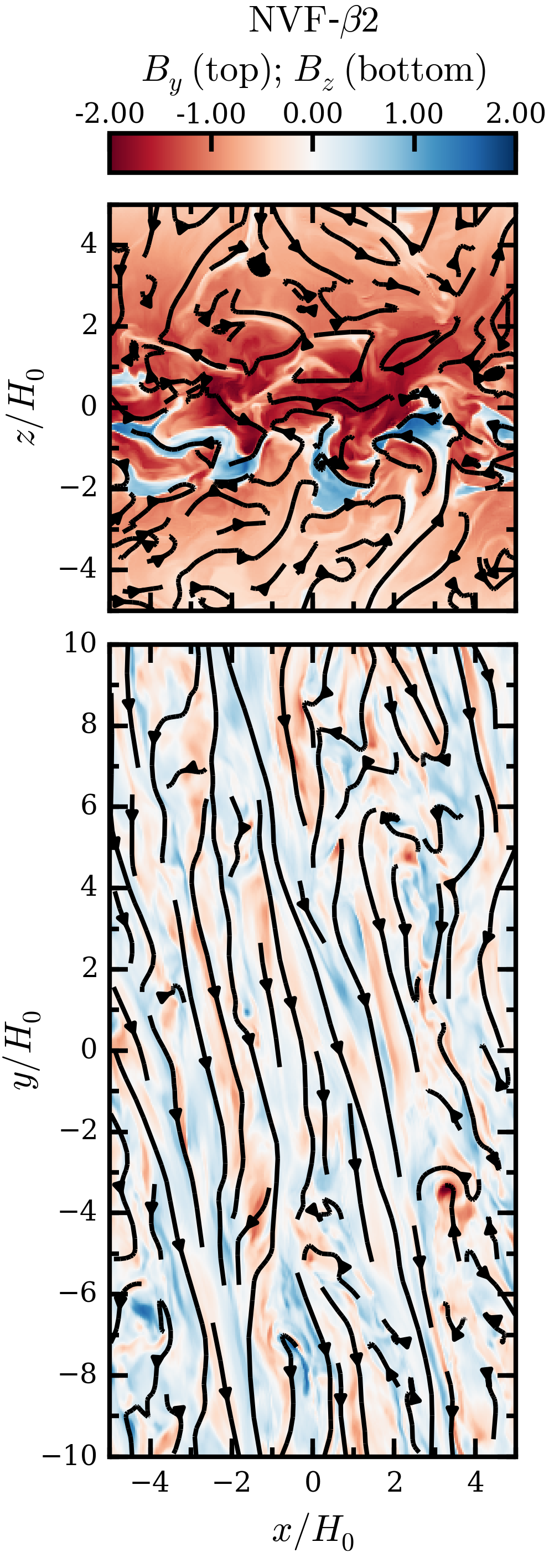}
  \caption{Magnetic field structure for simulations ZNVF-O ({\it left panels}), ZNVF-P ({\it center panels}), and NVF-$\beta2$ ({\it right panels}) at the snapshot in time $t = 100~{\rm orbits}$.  The {\it top panels} show streamlines for the poloidal magnetic field components and colours show the toroidal magnetic field strength for a slice through the $xz$-plane at $y = 0$.  The {\it bottom panels} show streamlines for the horizontal magnetic field components and colours show the vertical magnetic field strength for a slice through the $xy$-plane at $z = 0$.  In all cases, the toroidal magnetic field dominates and is well-organized at the disc mid-plane.  Notably, the toroidal magnetic field has large-scale coherence and is most highly organized for the strongly magnetized disc simulation NVF-$\beta2$.}
  \label{fig:Bcomps}
\end{figure*}

Figure \ref{fig:Bcomps} shows the magnetic field structure at the snapshot in time $t=100~{\rm orbits}$.  In all cases, the toroidal magnetic field is the dominant component.  Small-scale turbulent magnetic structures exist in the vicinity of the disc mid-plane for the weakly magnetized simulations (ZNVF-O and ZNVF-P).  Compared to ZNVF-O, simulation ZNVF-P shows an abrupt transition from the highly-turbulent disc core regions into the magnetically dominated disc atmosphere.  The ordered toroidal magnetic field structure in the disc mid-plane is similar for both ZNVF-O and ZNVF-P.  As discussed in \citet{Salvesen2016a}, the strongly magnetized disc simulation NVF-$\beta2$ develops ribbon-like field structures near the disc mid-plane and the field is remarkably well-ordered at the mid-plane.

Our results do not support suggestions that there are non-MRI dynamo routes to sustaining strong magnetization in the absence of poloidal field. For zero net vertical flux simulations with vertical boundary conditions that enforce the confinement of toroidal magnetic flux, we do find a very strongly magnetized corona ($\widehat{\beta} < 0.1$ for $|z| > 2 H_{0}$).  However, even in this case the disc evolves toward a weakly magnetized core ($\widehat{\beta}^{\rm mid} \simeq 32$).  For more physically meaningful boundary conditions that allow mass and magnetic flux to buoyantly escape, the initially strong toroidal flux is quickly expelled and the disc settles into a weakly magnetized state ($\widehat{\beta}^{\rm mid} \simeq 45$) that appears identical to similar simulations initialized with much weaker toroidal fields. We find no local limit in which a strongly magnetized disc (i.e., $\widehat{\beta}^{\rm mid} \lesssim 1$) can be sustained given an initial configuration with zero net vertical flux and outflow boundary conditions.

\section{Discussion and Conclusions}
\label{sec:discconc}
In this Letter we have argued that a strong poloidal magnetic flux threading the accretion disc is necessary in order to generate and sustain a magnetically dominated state. This conclusion differs from that of \citet{JohansenLevin2008}, who observed a strong magnetization ($\beta^{\rm mid} \simeq 2$) in zero net vertical flux simulations that was sustained for at least a moderate duration (20-30 orbits). It appears likely that the difference between our results is a consequence of the different vertical boundary conditions. Although we never find a strongly magnetized disc mid-plane, we do find strong magnetization at quite small heights above the mid-plane if we adopt periodic boundary conditions that --- like those used by \citet{JohansenLevin2008} --- confine the toroidal flux within the simulation domain\footnote{For technical reasons, we have not been able to run a case with {\em identical} boundary conditions to those used by \citet{JohansenLevin2008}, and hence some differences are expected between our run ZNVF-P and theirs.}. In the more physically-motivated situation where magnetic flux can escape, the initially strong flux is buoyantly expelled and the disc settles into a weakly magnetized state characteristic of local disc simulations with zero/weak net vertical flux. As we have noted previously \citep{Salvesen2016a}, the properties of the MRI dynamo do vary when the disc becomes strongly magnetized, with the characteristic reversals of toroidal field becoming increasingly infrequent and possibly entirely absent. Given this, it is possible that the Parker instability \citep{Parker1966, Shu1974} plays a role in the dynamo process in highly magnetized discs, as envisaged by \citet{ToutPringle1992} and \citet{JohansenLevin2008}. We suggest, however, that a net poloidal flux is a pre-requisite for any dynamo that leads to strongly magnetized accretion discs.

Even without a zero net vertical flux route to strong magnetization, there are plausible pathways to forming strongly magnetized discs via poloidal fields. The requisite field threading the disc may be generated {\it in situ} and/or externally.  In the {\it in situ} scenario, random small-scale field generated by dynamo activity within local patches of the disc could combine coherently to form a large-scale poloidal magnetic field \citep{ToutPringle1996, King2004}. It is known that this process generates poloidal fields of sufficient strength to locally modify the MRI \citep{Sorathia2010, Beckwith2011}, but whether the ultimate outcome of such an inverse cascade is sufficient to form a strongly magnetized disc is unclear.  In the external field scenario, a pervading background magnetic field could provide the poloidal flux.  This ``fossil field'' could be an interstellar magnetic field or a galactic field in the context of accretion discs in X-ray binaries and around supermassive black holes, respectively.  Regardless of the origin of the net poloidal flux, the efficiency of radial transport of flux through the accretion disc depends upon the effective magnetic diffusivity and the geometric thickness of the disc \citep{Lubow1994, GuiletOgilvie2012}. Given favorable conditions for accumulation of poloidal magnetic flux, a strongly magnetized disc would necessarily follow. In X-ray binaries, where there is strong evidence that the disc alternates between geometrically thick and thin configurations near the black hole, changes in the 
transport efficiency could lead to cyclic accumulation and loss of flux \citep{BegelmanArmitage2014}. 

Direct measurement of disc magnetic fields remains difficult. Numerous accretion disc phenomena, however, appear to be sensitive to the degree of magnetization, and further study of these may allow for useful constraints. Examples include the variability properties of radiation pressure dominated discs, which are predicted to be viscously/thermally stable when magnetic pressure-dominated \citep{BegelmanPringle2007}, unlike standard disc models without magnetic pressure support \citep{LightmanEardley1974, ShakuraSunyaev1976}, and the size of discs in active galactic nuclei, which are predicted to be more stable against fragmentation into stars when strongly magnetized \citep{Pariev2003, BegelmanPringle2007, Gaburov2012} than otherwise \citep{ShlosmanBegelman1987}.  From a more directly observational perspective, vertical magnetic pressure support acts to harden the emergent accretion disc spectrum \citep[e.g.,][]{Blaes2006}. For extremely magnetized discs, \citet{BegelmanPringle2007} predict dramatic spectral hardening with a colour correction factor of $f_{\rm col} \sim 5$. We have previously argued that the evolution of the disc spectral component during black hole X-ray binary state transitions can be explained by a variable $f_{\rm col}$, arising from changes in the disc vertical structure \citep{Salvesen2013}, and we are currently working to quantify the degeneracies between $f_{\rm col}$ and black hole spin determined via the continuum fitting technique (Salvesen et al., in preparation).  

For a geometrically thin disc supporting an MRI-dynamo, buoyant fields escape vertically on a time scale \citep[comparable to the free-fall time scale;][]{Begelman2015, Salvesen2016a} that is short compared to the time scale for viscous radial advection of $B_\phi$. The local approximation, used by both \citet{JohansenLevin2008} and ourselves, should then be appropriate. For somewhat thicker discs, long-duration global accretion disc simulations with varying levels of background poloidal magnetic flux are needed. Promising initial studies of strongly magnetized discs in the global limit have already been completed \citep{Machida2000, Gaburov2012, Sadowski2016}. Currently, however, the presence of a net poloidal magnetic flux is the only demonstrated way to generate and maintain an accretion disc with a dynamically important magnetic field (i.e., $\beta^{\rm mid} \sim 1$).  Future work on disc accretion physics in this strongly magnetized regime may help decipher the enigmatic phenomena associated with accretion discs in X-ray binaries and galactic nuclei.

\section*{Acknowledgments}
We thank the referee, Anders Johansen, for his careful reading and constructive comments.  GS acknowledges support through the NASA Earth and Space Science Graduate Fellowship program.  PJA acknowledges support from NASA under Astrophysics Theory Program awards NNX11AE12G and NNX14AB42G, and from the NSF under award AST-1313021.  J.B.S.'s support was provided in part under contract with the California Institute of Technology (Caltech) and the Jet Propulsion Laboratory (JPL) funded by NASA through the Sagan Fellowship Program executed by the NASA Exoplanet Science Institute.  MCB acknowledges support from NSF grant AST-1411879.  This work used the \texttt{Janus} supercomputer, which is supported by the National Science Foundation (award number CNS-0821794) and the University of Colorado Boulder.  The \texttt{Janus} supercomputer is a joint effort of the University of Colorado Boulder, the University of Colorado Denver, and the National Center for Atmospheric Research.

\bibliographystyle{mn2e}
\bibliography{/Users/salvesen/ms/bib/salvesen}

\label{lastpage}
\end{document}